\documentclass[fleqn,usenatbib]{mnras}
\usepackage{newtxtext,newtxmath}

\usepackage[T1]{fontenc}
\usepackage{ae,aecompl}

\usepackage{graphicx}	
\usepackage{times}

\title[The H-D limit revisited]{The luminosities of cool supergiants in the Magellanic Clouds, and the Humphreys-Davidson limit revisited}

\author[B. Davies, P.A. Crowther \& E.R. Beasor]{
Ben Davies,$^{1}$\thanks{}, Paul A.\ Crowther$^{2}$ and Emma R.\ Beasor$^{1}$
\\
$^{1}$Astrophysics Research Institute, Liverpool John Moores 
University, Liverpool Science Park ic2, \\ 146 Brownlow Hill, Liverpool, L3 5RF, UK\\
$^{2}$Dept of Physics \& Astronomy, University of Sheffield, Hounsfield Rd, Sheffield S3 7RH, UK\\
}

\date{Accepted XXX. Received YYY; in original form ZZZ}

\pubyear{2017}
\hypersetup{draft}
\begin{document}
\label{firstpage}
\pagerange{\pageref{firstpage}--\pageref{lastpage}}
\maketitle
      
\begin{abstract}
The empirical upper luminosity boundary $L_{\rm max}$ of cool supergiants, often referred to as the Humphreys-Davidson limit, is thought to encode information on the general mass-loss behaviour of massive stars. Further, it delineates the boundary at which single stars will end their lives stripped of their hydrogen-rich envelope, which in turn is a key factor in the relative rates of Type-II to Type-Ibc supernovae from single star channels. In this paper we have revisited the issue of $L_{\rm max}$ by studying the luminosity distributions of cool supergiants (SGs) in the Large and Small Magellanic Clouds (LMC/SMC). We assemble samples of cool SGs in each galaxy which are highly-complete above $\log L/L_{\odot}$=5.0, and determine their spectral energy distributions from the optical to the mid-infrared using modern multi-wavelength survey data. We show that in both cases $L_{\rm max}$ appears to be lower than previously quoted, and is in the region of $\log L/L_{\odot}$=5.5. There is no evidence for $L_{\rm max}$ being higher in the SMC than in the LMC, as would be expected if metallicity-dependent winds were the dominant factor in the stripping of stellar envelopes. We also show that $L_{\rm max}$ aligns with the lowest luminosity of single nitrogen-rich Wolf-Rayet stars, indicating of a change in evolutionary sequence for stars above a critical mass. From population synthesis analysis we show that the Geneva evolutionary models greatly over-predict the numbers of cool SGs in the SMC. We also argue that the trend of earlier average spectral types of cool SGs in lower metallicity environments represents a genuine shift to hotter temperatures. Finally, we use our new bolometric luminosity measurements to provide updated bolometric corrections for cool supergiants. 
\end{abstract}

\begin{keywords}
stars: massive -- stars: evolution -- supergiants
\end{keywords}

\def\ga{\mathrel{\hbox{\rlap{\hbox{\lower4pt\hbox{$\sim$}}}\hbox{$>$}}}}
\def\la{\mathrel{\hbox{\rlap{\hbox{\lower4pt\hbox{$\sim$}}}\hbox{$<$}}}}
\def\msunyr{M\mbox{$_{\normalsize\odot}$}\rm{yr}$^{-1}$}
\def\msun{$M$\mbox{$_{\normalsize\odot}$}}
\def\zsun{$Z$\mbox{$_{\normalsize\odot}$}}
\def\rsun{$R$\mbox{$_{\normalsize\odot}$}}
\def\minit{$M_{\rm init}$}
\def\lsun{$L$\mbox{$_{\normalsize\odot}$}}
\def\mdot{$\dot{M}$}
\def\mdotdj{$\dot{M}_{\rm dJ}$}
\def\lbol{$L$\mbox{$_{\rm bol}$}}
\def\kms{\,km~s$^{-1}$}
\def\EW{$W_{\lambda}$}
\def\arcsec{$^{\prime \prime}$}
\def\arcmin{$^{\prime}$}
\def\teff{$T_{\rm eff}$}
\def\Teff{$T_{\rm eff}$}
\def\logg{$\log g$}
\def\logz{$\log Z$}
\def\logl{$\log (L/L_\odot)$}
\def\vdisp{$v_{\rm disp}$}
\def\bcv{{\it BC$_V$}}
\def\bci{{\it BC$_I$}}
\def\bck{{\it BC$_K$}}
\def\lmax{$L_{\rm max}$}
\def\um{$\mu$m}
\def\chisq{$\chi^{2}$}
\def\AV{$A_{V}$}
\def\hminus{H$^{-}$}
\def\Hminus{H$^{-}$}
\def\ebmv{$E(B-V)$}
\def\mdyn{$M_{\rm dyn}$}
\def\mphot{$M_{\rm phot}$}
\def\cnterm{[C/N]$_{\rm term}$}
\newcommand{\fig}[1]{Fig.\ \ref{#1}}
\newcommand{\Fig}[1]{Figure \ref{#1}}
\newcommand{\newtext}[1]{{#1}}
\newcommand{\nntext}[1]{{#1}}


\section{Introduction}
Models of stellar evolution predict that stars with initial masses $M_{\rm init} \geq 8 M_{\odot}$ should swell up to become cool supergiants (SGs) when they leave the main-sequence. However, it has long since been established that there is an upper luminosity limit \lmax\ above which no cool SGs are observed \citep{Stothers69,Sandage-Tammann74}, now commonly referred to as the Humphreys-Davidson (H-D) limit \citep{Humphreys-Davidson79}. The existence of this limit implies that the highest mass stars do not evolve to the cool side of the Hertzsprung-Russell (H-R) diagram and instead remain more compact, ending their lives as either blue hypergiants or Wolf-Rayet stars. 

The common interpretation of this luminosity limit is that it is caused by mass-loss, either via a smooth wind, or by episodic Luminous Blue Variable (LBV) type eruptions: the more massive the star, the stronger the mass-loss, resulting in a larger fraction of the star's initial mass being lost prior to core-collapse supernova (ccSN). Above some initial mass threshold, the entire H-rich envelope can be lost before the star can evolve to the cool side of the H-R diagram, causing it to evolve directly to the Wolf-Rayet (WR) phase. Just below this mass limit, stars are expected to have a brief cool SG phase before becoming a WR \citep[e.g.][]{Stothers-Chin79,Chiosi-Maeder86}. Therefore, \lmax\ is sensitive to the mass-loss rates of stars integrated over their lifetimes. 

The most luminous Red Supergiants (RSGs) identified by \citet{Humphreys-Davidson79} in the Milky Way and Large Magellanic Cloud (LMC) were inferred to have $\log L/L_{\odot}$ = 5.74 and 5.66, respectively, interpreted as reflecting a genuine limit at $\log L/L_{\odot} = 5.8\pm$0.1. These measurements of \lmax\ relied upon assumed optical bolometric corrections for RSGs, uncertain distances to the Galactic cool supergiants\newtext{, an outdated distance modulus to the LMC}, and a selective sample of optically-bright stars. Hence, dust-enshrouded cool hypergiants \citep[e.g.][]{vL05} would have been missed from their optical study, while those with moderate circumstellar extinction may have had their luminosities underestimated. More recently, determination of bolometric luminosities \lbol\ for the large samples of cool SGs in the MCs has been attempted from atmospheric model fitting \citep[e.g.][ hereafter L06]{Levesque06}, from which visible and near-infrared (IR) bolometric corrections are obtained \citep{Neugent12}. However, determining \lbol\ in this way is problematic, due to the presence of circumstellar dust and/or the inherent deficiencies in 1-D atmospheric models when used to model highly anisotropic stars \citep[][]{Levesque09,rsgteff}. 

By focusing exclusively on the Magellanic Clouds (MCs) it is possible to negate the impact of uncertain distances and high foreground extinction. Further, by adding in near- and mid-IR photometry, we can compensate for circumstellar extinction under the assumption that the flux lost at short wavelengths is re-radiated in the mid-IR. We can then obtain bolometric luminosities by directly integrating under the spectral energy distributions. This is possible thanks to extensive photometric surveys of luminous stars in the MCs that have been conducted within recent decades, both visually \citep{Massey02,Zaritsky02,Zaritsky04} and in the near- and mid-IR \citep{SAGE-LMC,SAGE-SMC,WISE}. These surveys have been mined in attempts to obtain statistically complete samples of cool supergiants, with hundreds of candidates subsequently being confirmed with spectroscopic follow-ups \citep[e.g.][]{Massey-Olsen03,Neugent12,Gonzalez-Fernandez15}.



In this paper we take a fresh look at the luminosity distribution of cool supergiants in the MCs, with a particular focus on \lmax. In Sect.\ \ref{sec:obs} we describe the input catalogues we employ to compile our list of targets, and the multi-wavelength photometry we use to determine model-independent luminosities for each target. In Sect. \ref{sec:ldist} we then construct model-independent luminosity distributions of cool supergiants for both the LMC and SMC. We revisit the issue of \lmax, compare the luminosity distribution of cool SGs to that of WRs, and to the predictions of evolutionary models. We conclude in Sect.\ \ref{sec:conc}. In the Appendix we also provide a reappraisal of the bolometric corrections of cool supergiants.

\begin{figure*}
\begin{center}
\includegraphics[width=7.5cm]{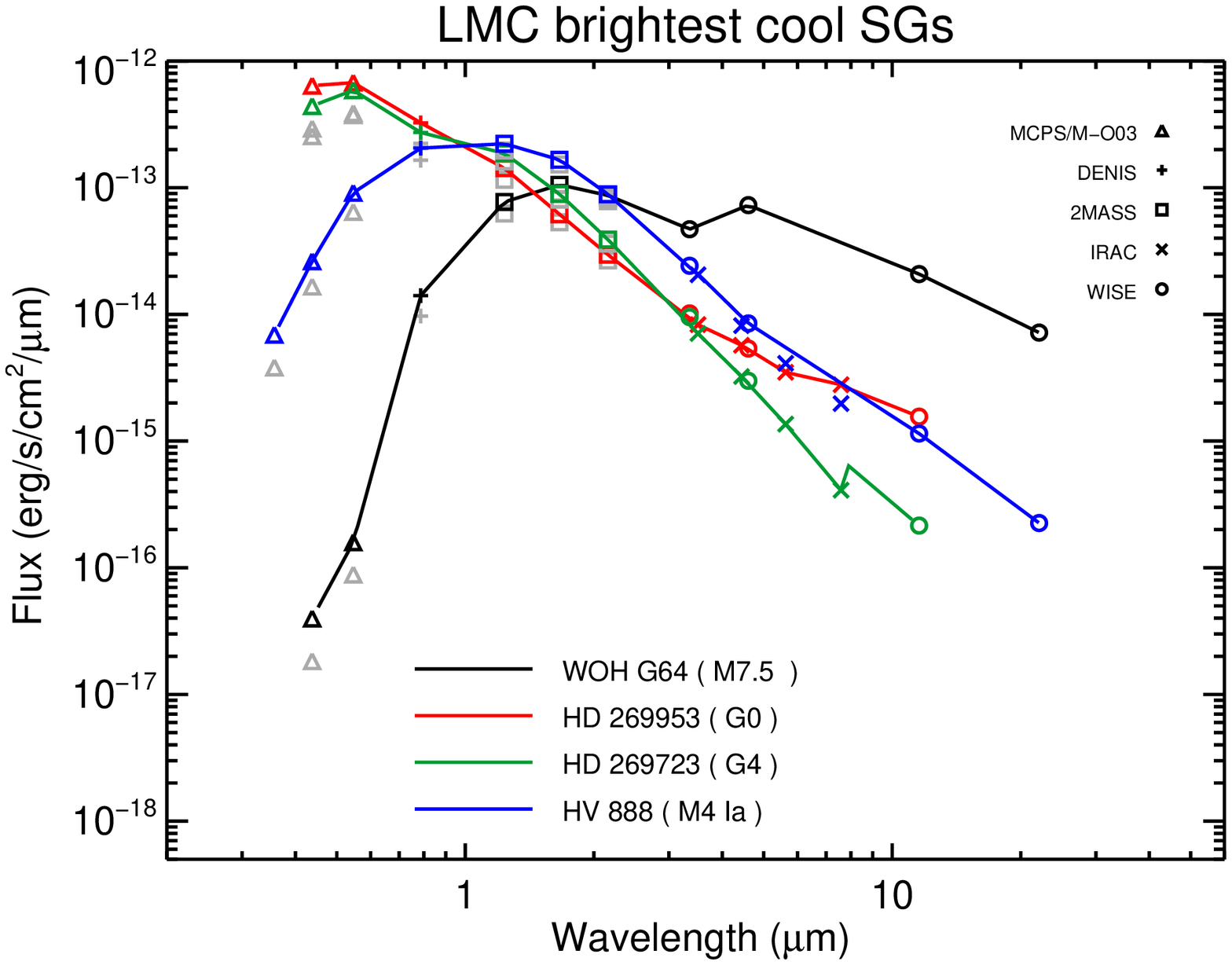}
\includegraphics[width=7.5cm]{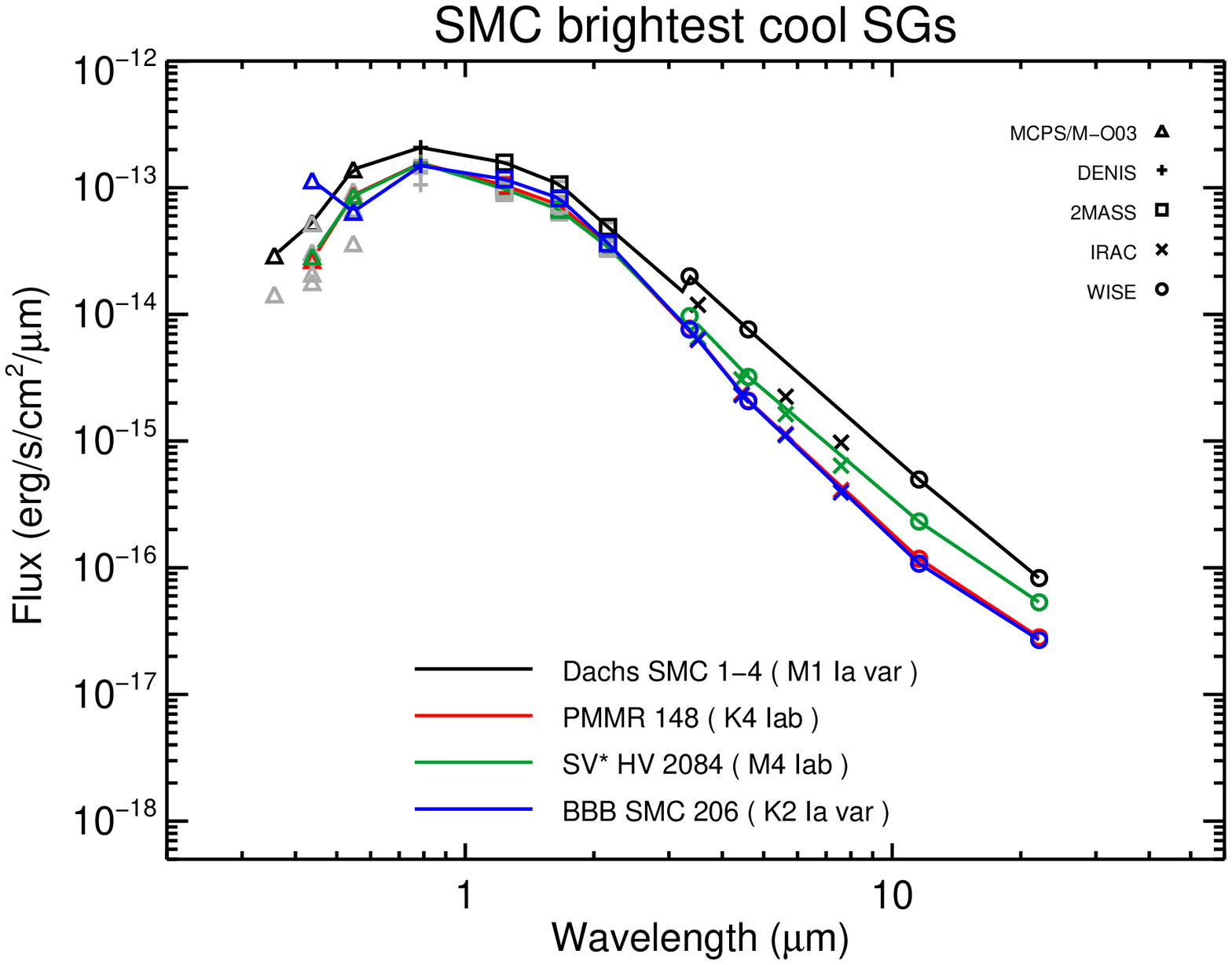}
\caption{Observed spectral energy distributions of the 3 most luminous stars in each galaxy.  The different symbols indicate the source of the photometry, as described by the legend in the upper left of each panel. The grey symbols show the photometry prior to correction for reddening.}
\label{fig:sed}
\end{center}
\end{figure*}

\section{Observational sample} \label{sec:obs}

\subsection{Input catalogues}
To compile a list of cool SGs in each of the Magellanic Clouds we have pooled data from several input catalogues, each of which uses a different technique to identify candidate objects. We do this so as to be as complete as possible; dust-enshrouded stars which may be too faint at visible wavelengths to be found in optical surveys may instead show up in mid-IR catalogues. The earliest spectral type we consider is G0, since we wish to separate cool stars from LBV-like objects which may temporarily evolve from the blue to F-types \citep[e.g. R71,][]{Mehner13,Mehner17}. Below, we describe the catalogues we have targeted, acknowledging that there is a large degree of overlap between these catalogues. 

\begin{itemize}
\item For optically-selected targets, we used \citet{Elias85}, \citet{Levesque06,Levesque07}, and \citet{Neugent10,Neugent12} which built upon earlier work of \citet{Humphreys79-SMC,Humphreys79-LMC} and \citet{Massey-Olsen03}. In each study  candidates were selected on the basis of optical colours and brightnesses, and were confirmed with follow-up spectroscopy. 
\item For near-infrared (IR) bright sources, we used the catalogue of \citet{Gonzalez-Fernandez15}. Targets were selected on the basis of near-IR photometry from 2MASS, with spectral types confirmed from follow-up optical spectroscopy.
\item For targets bright in the mid-IR, we used the catalogues compiled from the {\it Spitzer} SAGE survey by \citet{Bonanos09,Bonanos10}. Objects were classified on the basis of their mid-IR colours, a technique calibrated by stars in the two MCs with known spectral types from \citet{Humphreys79-SMC,Humphreys79-LMC} and \citet{Massey-Olsen03}.
\item In addition to the above, we also used the LMC study of \citet{Buchanan06} which selected bright 8\um\ sources from {\it Spitzer/IRAC} and classified them on the basis of {\it Spitzer/IRS} mid-IR spectroscopy. From this catalogue we selected those sources confirmed to belong to the LMC, and which had O-rich signatures in their spectra. 
\item \newtext{Finally, we took the samples of dusty and/or maser-emitting RSGs from \citet{vL05}, \citet{Goldman17} and \citet{Goldman18}. These stars are thought to have thick dusty envelopes, and so are often very faint in the optical, but are spectroscopically confirmed RSGs. }
\end{itemize} 

For each source in our master database, we then collate (where available) {\it UBV} photometry from the Magellanic Clouds Photometric Survey \citep[MCPS, ][]{Zaritsky02,Zaritsky04}, {\it BV} photometry from \citet[][ hereafter M-O03]{Massey-Olsen03}, $I$-band photometry from {\it DENIS} \citep{DENIS}, {\it JHK} photometry from 2MASS \citep{2MASS}, and mid-IR photometry from {\it Spitzer}/IRAC \citep{SAGE-LMC,SAGE-SMC} and WISE \citep{WISE}. To aid in vetting our source list of foreground interlopers, we also search for proper motion measurements from {\it Hipparcos} \citep{HIPPARCOS}, and radial velocity measurements and luminosity classifications from \citet{Gonzalez-Fernandez15}. Targets are rejected from the catalogue if they have any of the following:

\begin{itemize}
\item Heliocentric radial velocities $v_{\rm hel}$ less than 70\kms\ below the average for their putative host galaxy, i.e. $v_{\rm hel} < 80$\kms\ and $v_{\rm hel} < 200$\kms\ for the SMC and LMC respectively, based on the results of \citet{Gonzalez-Fernandez15}.
\item Proper motions greater than 1mas/yr, from {\it Hipparcos} \citep{HIPPARCOS}.
\item Luminosity classes of II or fainter, based on \citet{Gonzalez-Fernandez15}.
\end{itemize}

The foreground extinction to each star was determined from the extinction maps of \citet{Zaritsky02,Zaritsky04}, which were themselves constructed from the apparent colours of hot stars. For each star in our catalogue, we took the visual extinction \AV\ to be the median of that at the star's position and the neighbouring 8 pixels (corresponding to a radius of 1\arcmin), with the error taken to be the standard deviation. We then dereddened the star's photometry according to the extinction law in \citet{Gordon03} appropriate for the star's host galaxy. 

\begin{table*}
\caption{Name, position, luminosity and extinction of the 20 most luminous cool SGs in each of the LMC and SMC. In this table we provide for each star the SIMBAD designation, as well as those from \citet{Massey02} and \citet{Gonzalez-Fernandez15} where available. Where stars are known to be variable, we list the minimum and maximum known spectral types and luminosity classes. Full observational information on all stars in this study (over 300 per galaxy), including photometry from the $U$-band to 70\um, is available on-line at XXXXXX.}
\begin{center}
\begin{tabular}{lccccccc}
\hline \hline
SIMBAD Name & $[$M2002] & $[$GDN2015$]$ & Spec Type & RA DEC (J2000) & $\log(L/L_{\odot})$ & $A_V$ \\
\hline
{\it LMC} \\
                WOH G064 &            &        &           M7.5  &  04 55 10.48  -68 20 29.8 & 5.77 $\pm$ 0.04 & 0.72 $\pm$ 0.12 \\
              HD 269953 &            &        &              G0 &  05 40 12.18  -69 40 05.0 & 5.50 $\pm$ 0.04 & 0.72 $\pm$ 0.15 \\
              HD 269723 &            &        &              G4 &  05 32 24.96  -67 41 53.7 & 5.48 $\pm$ 0.08 & 0.51 $\pm$ 0.34 \\
                 HV 888 &            &        &           M4 Ia &  05 04 14.14  -67 16 14.4 & 5.48 $\pm$ 0.04 & 0.42 $\pm$ 0.19 \\
            SV* HV 2450 &            &        &           M2 Ia &  05 19 53.26  -68 04 03.8 & 5.45 $\pm$ 0.04 & 0.58 $\pm$ 0.18 \\
             SP77 46-44 & LMC 145013 &        &      M2.5 Ia-Ib &  05 29 42.21  -68 57 17.4 & 5.40 $\pm$ 0.02 & 0.12 $\pm$ 0.08 \\
            SV* HV 5618 & LMC 071357 &        &            M1 I &  05 07 05.66  -70 32 44.0 & 5.38 $\pm$ 0.04 & 0.32 $\pm$ 0.21 \\
             SP77 31-16 &            &        &                 &  04 54 36.84  -69 20 22.1 & 5.35 $\pm$ 0.03 & 0.56 $\pm$ 0.15 \\
            LI-LMC 1100 &            &        &                 &  05 27 40.78 -69 08 05.7 & 5.34 $\pm$ 0.00 & 0.50 $\pm$ 0.10 \\
          $[$MG73$]$ 46 &            &        &                 &  05 35 55.23  -69 09 59.5 & 5.34 $\pm$ 0.04 & 1.10 $\pm$ 0.21 \\
 $[$M2002$]$ LMC 165543 & LMC 165543 &        &           G1 Ia &  05 36 26.79  -69 23 51.4 & 5.33 $\pm$ 0.04 & 0.49 $\pm$ 0.19 \\
   $[$GDN2015$]$ LMC252 &            & LMC252 &        M0 Ia-Ib &  05 39 32.34  -69 34 50.1 & 5.30 $\pm$ 0.04 & 0.89 $\pm$ 0.24 \\
 $[$M2002$]$ LMC 144217 & LMC 144217 &        &           M3 Ia &  05 29 27.58  -69 08 50.3 & 5.30 $\pm$ 0.03 & 0.51 $\pm$ 0.14 \\
                HV 2561 & LMC 141430 &        &           M2 Ia &  05 28 28.86  -68 07 07.9 & 5.29 $\pm$ 0.05 & 0.76 $\pm$ 0.27 \\
 $[$M2002$]$ LMC 136042 & LMC 136042 &        &        M4 Ia-Ib &  05 26 34.80  -68 51 40.0 & 5.27 $\pm$ 0.03 & 0.34 $\pm$ 0.16 \\
                 HV 916 &            &        &           M3 Ia &  05 14 49.72  -67 27 19.7 & 5.27 $\pm$ 0.04 & 0.76 $\pm$ 0.23 \\
    $[$GDN2015$]$ LMC45 &            & LMC45  &           M3 Ia &  05 26 23.54  -69 52 25.8 & 5.27 $\pm$ 0.04 & 0.22 $\pm$ 0.16 \\
             LI-LMC 183 & LMC 023095 &        &             M2  &  04 55 03.07  -69 29 12.8 & 5.24 $\pm$ 0.03 & 0.49 $\pm$ 0.18 \\
            SV* HV 2595 & LMC 147199 &        &            M4 I &  05 30 20.94  -67 20 05.4 & 5.23 $\pm$ 0.03 & 0.30 $\pm$ 0.16 \\
              WOH S 229 & LMC 113364 &        &            M1 I &  05 19 03.26  -69 39 55.3 & 5.23 $\pm$ 0.03 & 0.42 $\pm$ 0.12\smallskip \\

{\it SMC} \\
          Dachs SMC 1-4 & SMC 018592 &        &   M0.5-M3 Ia-Ib &  00 51 03.86  -72 43 17.6 & 5.55 $\pm$ 0.01 & 0.48 $\pm$ 0.05 \\
               PMMR 148 & SMC 056389 &        &          K4 Iab &  01 03 27.64  -72 52 09.6 & 5.39 $\pm$ 0.03 & 0.36 $\pm$ 0.14 \\
            SV* HV 2084 & SMC 069886 & SMC400 &          M4 Iab &  01 09 38.24  -73 20 02.4 & 5.35 $\pm$ 0.02 & 0.28 $\pm$ 0.09 \\
   $[$GDN2015$]$ SMC354 &            & SMC354 &           G1 Ib &  01 03 53.87  -72 45 15.0 & 5.34 $\pm$ 0.05 & 0.53 $\pm$ 0.21 \\
                PMMR 37 & SMC 018136 &        &      K4.5 Ia-Ib &  00 50 56.09  -72 15 06.1 & 5.33 $\pm$ 0.03 & 0.36 $\pm$ 0.14 \\
           LHA 115-S 30 & SMC 049478 &        &        K5 Ia-Ib &  01 00 41.51  -72 10 37.1 & 5.27 $\pm$ 0.04 & 0.40 $\pm$ 0.18 \\
           SV* HV 11423 & SMC 050028 &        &          M0Iab  &  01 00 55.20  -71 37 52.9 & 5.25 $\pm$ 0.02 & 0.27 $\pm$ 0.11 \\
            BBB SMC 206 & SMC 010889 &        &     K2-K5 Ia-Ib &  00 48 27.02  -73 12 12.3 & 5.25 $\pm$ 0.02 & 0.70 $\pm$ 0.07 \\
            SV* HV 1475 & SMC 013472 &        &   G5.5-M0 Ia-Ib &  00 49 24.55  -73 18 13.6 & 5.24 $\pm$ 0.02 & 0.59 $\pm$ 0.09 \\
         Dachs SMC 2-37 & SMC 059803 &        &     G7-K3 Ia-Ib &  01 04 38.21  -72 01 27.0 & 5.24 $\pm$ 0.03 & 0.27 $\pm$ 0.18 \\
                 PMMR 9 & SMC 005092 &        &      M1.5 Ia-Ib &  00 45 04.57  -73 05 27.7 & 5.23 $\pm$ 0.03 & 0.79 $\pm$ 0.13 \\
  $[$M2002$]$ SMC 64663 & SMC 064663 &        &   G6-K3.5 Ia-Ib &  01 06 47.67  -72 16 11.8 & 5.21 $\pm$ 0.03 & 0.42 $\pm$ 0.15 \\
                PMMR 52 & SMC 025888 &        &      K3.5 Ia-Ib &  00 53 09.12  -73 04 03.8 & 5.21 $\pm$ 0.02 & 0.46 $\pm$ 0.05 \\
                LIN 235 &            &        &     K1-K4 Ia-Ib &  00 53 08.95  -72 29 38.6 & 5.20 $\pm$ 0.02 & 0.54 $\pm$ 0.10 \\
            BBB SMC 138 & SMC 012322 &        &   K3-M1.5 Ia-Ib &  00 49 00.35  -72 59 35.9 & 5.19 $\pm$ 0.03 & 0.72 $\pm$ 0.14 \\
                PMMR 70 & SMC 030616 &        &   K0-K2.5 Ia-Ib &  00 54 35.90  -72 34 14.4 & 5.18 $\pm$ 0.02 & 0.54 $\pm$ 0.11 \\
                PMMR 41 & SMC 020133 &        &   K3-M1.5 Ia-Ib &  00 51 29.68  -73 10 44.2 & 5.18 $\pm$ 0.02 & 0.54 $\pm$ 0.08 \\
    $[$GDN2015$]$ SMC54 &            & SMC54  &                 &  00 42 17.14  -74 06 15.3 & 5.16 $\pm$ 0.08 & 0.18 $\pm$ 0.38 \\
                PMMR 62 &            &        &   K3-M1.5 Ia-Ib &  00 53 47.94  -72 02 09.5 & 5.16 $\pm$ 0.05 & 0.34 $\pm$ 0.22 \\
               PMMR 105 & SMC 047757 &        &        K0 Ia-Ib &  01 00 00.58  -72 19 40.4 & 5.14 $\pm$ 0.02 & 0.45 $\pm$ 0.08\smallskip \\
            \hline
\end{tabular}
\end{center}
\label{tab:20}
\end{table*}%

\subsection{Determining bolometric luminosities}
The dereddened photometry was first converted to fluxes using the filter profile information made available by the SVO Filter Profile Service\footnote{{\tt http://svo2.cab.inta-csic.es/theory/fps/}}. The spectral energy distributions (SEDs) were resampled onto a logarithmically-spaced wavelength axis using the {\tt spline} function in IDL, before being integrated using the IDL function {\tt int\_tabulated} to find their apparent bolometric luminosities. Absolute luminosities were determined from the distance moduli of 18.49 and 18.95 for the LMC and SMC respectively \citep{Pietrzynski13,Graczyk14}. Example plots of SEDs for the brightest sources in each galaxy are shown in \fig{fig:sed}.

In our study we have {\it not} corrected for circumstellar extinction, which is well-known to exist around many cool SGs \citep[e.g.][]{k-w98,vL05,dewit08}. Instead, we have made the assumptions that any flux lost to absorption by circumstellar material is re-radiated in the mid-IR, \newtext{and that the radiation is isotropic (the latter assumption is discussed further in Sect.\ \ref{sec:comp})}. For all but a small number of stars in our sample, the contribution to the total luminosity by the flux at our longest wavelength data-point ({\it WISE-4}, 24\um) is very small. For objects which are bright at 24\um, we also add in the flux at 70\um\ from \citet{Jones17}. Even for the brightest, reddest star in our sample (WOH G64), the flux at wavelengths $>$24\um\ is negligible (see Sect.\ \ref{sec:ldist}).

At the opposite end of the spectrum, we have estimated the amount of flux emitted at short wavelengths by creating a black-body spectrum and matching it to the dereddened $U$- or $B$-band flux. For K and M types, we use a black-body \teff\ of 4000K, for G types 5000K, unless the object already has a specific \teff\ estimate in \citet{Neugent10} or \citet{Neugent12}. The amount of flux emitted at these wavelengths by cool SGs is again small, only a few $\times$0.01dex. This contribution rises to $\sim$0.1dex for the earliest spectral types in our sample. 

A summary of the observational data on the twenty most luminous cool SGs in each galaxy are listed in Table \ref{tab:20}. Full details on all stars in this work, including all photometry, can be found on-line at \newtext{XXXX (CDS, vizier placeholder)}. 

\begin{figure*}
\begin{center}
\includegraphics[width=7.5cm]{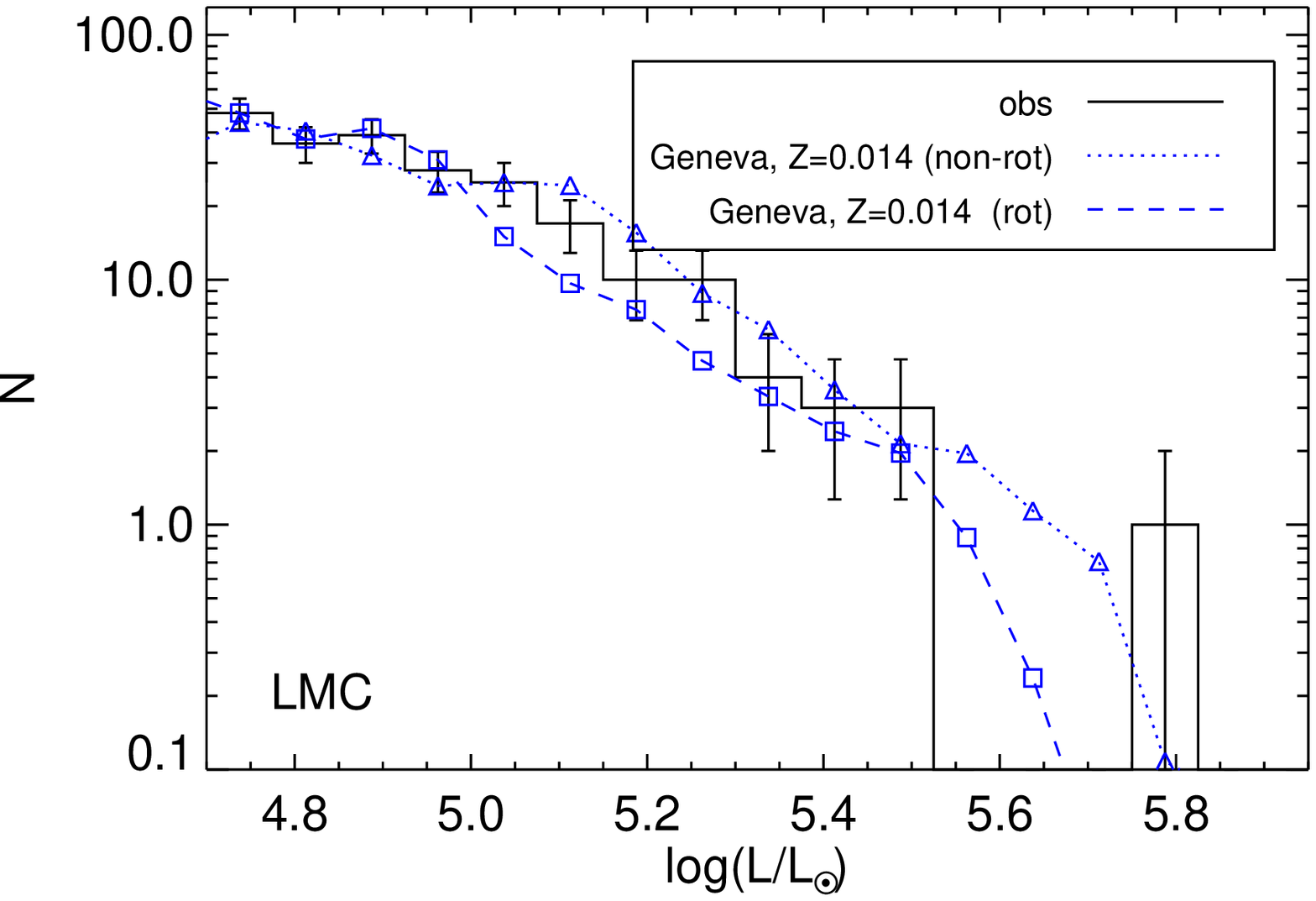}
\includegraphics[width=7.5cm]{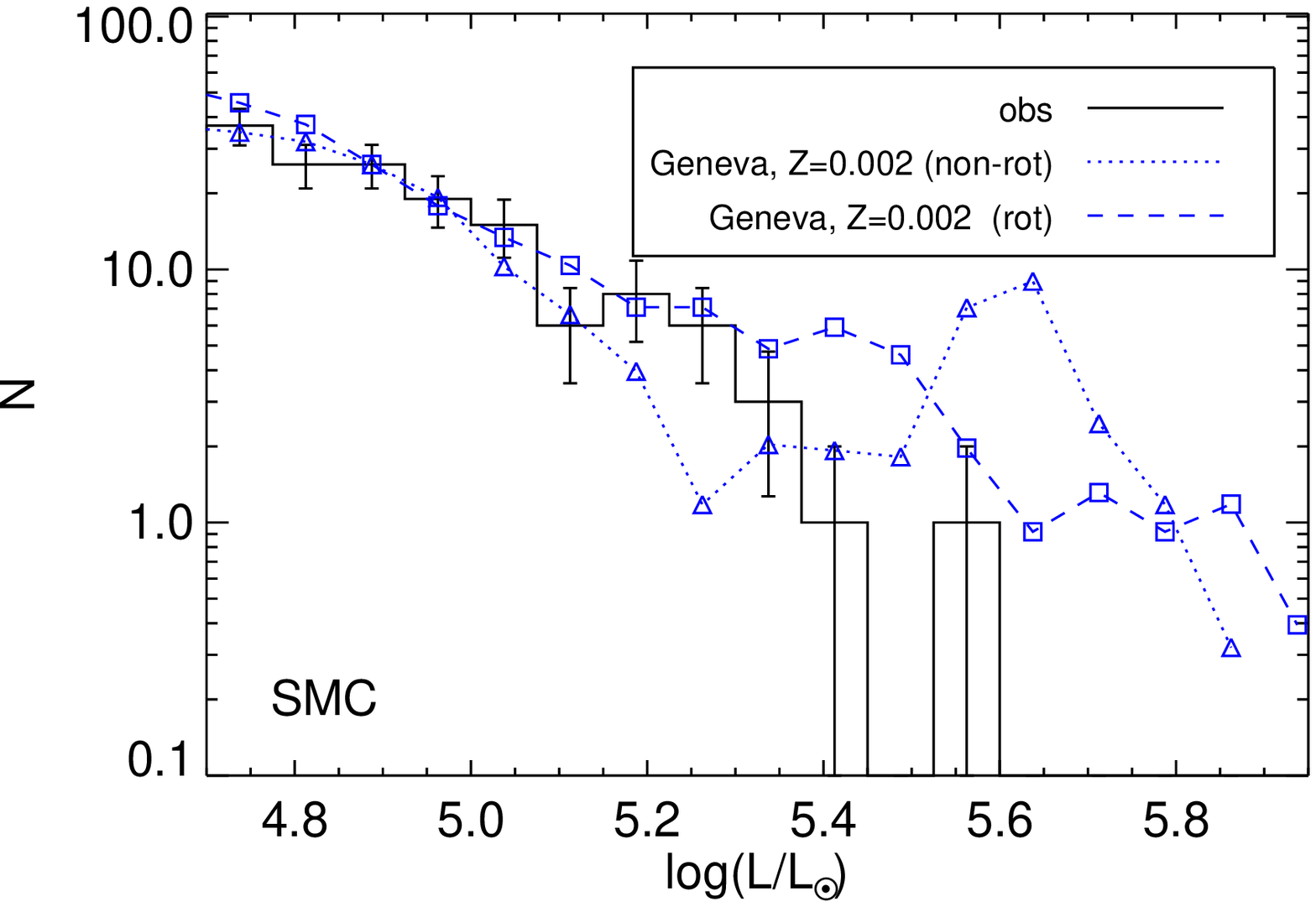}
\caption{Luminosity distributions for the LMC (left) and SMC (right). Overplotted are the Geneva model predictions for Solar \citep{Ekstrom12} and SMC-like \citep{Georgy13}. The model predictions have been normalised to fit the observations at \logl=4.7-5.2. }
\label{fig:ldist}
\end{center}
\end{figure*}

\section{Luminosity distributions and \lmax} \label{sec:ldist}
Histograms of the number of stars per luminosity bin for the two MCs are plotted in \fig{fig:ldist}. In both galaxies, we see what at first look appears to be a sharp cliff-edge to the luminosity distribution, with one more object $\sim$0.2dex brighter. There are three possible explanations for this edge: there is a genuine hard upper limit to the luminosities of cool SGs; the lifetime of the cool SG phase at high luminosities is very short; or it is caused by small number statistics at the upper end of the initial mass function. In the following sections we will argue that this cutoff is {\it not} an artefact of low number statistics. Though we cannot distinguish between a hard upper limit to $L$ and very short cool SG lifetimes above this limit, we will show that there is a genuine tension with evolutionary theory, particularly in the case of the SMC.

\subsection{Comparisons between the LMC and SMC} \label{sec:comp}
In the LMC, we see that there is an apparent truncation of the luminosity distribution at \logl=5.5. If we were to extrapolate beyond this limit at the gradient seen at \logl$>$5, we would expect to see $\approx$4 stars above \logl=5.5, whereas we see only one. This bright star is WOH~G64, with \logl=5.77. At this luminosity, stellar evolutionary models imply an initial mass of $\ga$40\msun, and an age of $\la$5Myr. Despite this young age, the star seems to be relatively isolated, with the closest markers of recent star formation such as ionized nebulae or other massive stars over 70\arcsec\ away \citep{Levesque09}. Further, this object is highly variable, with minimum-to-maximum variability ranging from 3mags at $B$ to 1.5mags at $I$ \citep[MACHO,][]{Soszyski09}. Further, the variability in the different wavebands appears to be positively correlated, implying a variable \lbol\ rather than in just colour. \nntext{Finally, we note that several authors have studied the SED of WOH~G64 and concluded that the circumstellar material cannot be spherically symmetric \citep{Roche93,vanLoon99,Ohnaka08,Goldman17}. In particular, Ohnaka et al.\ modelled the excess emission as originating in a dusty torus, which resulted in the star's luminosity being revised downwards to \logl=5.45, which would imply a much lower initial mass of around 25\msun}. Excluding WOH~G64, the five next most luminous stars cluster around 5.4$<$\logl$<$5.5, suggesting an upper luminosity limit for the LMC of \logl$=$5.5. 

In the SMC (\fig{fig:ldist}, right panel), the cliff-edge to the luminosity distribution occurs at a lower $L$, \logl=5.36. Again, from a simple extrapolation to higher luminosities we would expect to see $\approx$7 stars above this limit, rather than just the one star observed. The bright star is Dachs~SMC~1-4, a RSG with a luminosity of \logl=5.55. This star is detected in Gaia~DR1 as having a proper motion \citep{Gaia-DR1}, implying that it may be a foreground star \newtext{, though its radial velocity is consistent with the SMC}. However, the errors on this measurement are large, so until Gaia~DR2 we refrain from drawing any conclusions as to the nature of this star. Unlike WOH~G64, this star is only moderately variable, with a minimum-to-maximum amplitude of 0.25mags in $R$ \citep{Pojmanski02}. \newtext{It is therefore less easy to discount Dachs~SMC~1-4, and so it may indeed be representative of \lmax\ in the SMC. }

In each galaxy, we conservatively estimate the observed upper luminosity limit as being that of the second and third brightest stars, so as to insulate our conclusions from outliers and peculiar objects. Under this definition, the upper luminosity limits $L_{\rm max}$ for the two galaxies are \logl=5.4 and 5.5 for the SMC and LMC respectively. Before proceeding to study the predictions of stellar evolution models in detail, we first note that this behaviour of \lmax\ with metallicity goes opposite to the direction one would naively expect. At lower metallicity, mass-loss rates on the main-sequence (MS) should be lower, and so post-MS envelope masses should be higher. This would allow stars with higher masses to evolve to the RSG stage in lower metallicity environments. However, we see the opposite: \lmax\ is higher in the LMC, where the metallicity is roughly twice that of the SMC \citep{MCpaper}. 

To investigate whether the above result is an artefact of number statistics, we perform a simple numerical experiment in which we model the SMC simply as a scaled-down version of the LMC\footnote{The current star formation rate of the SMC
is $\sim$6 times lower than the LMC \citep{Kennicutt08} so their similar cool SG populations supports a significantly
higher ratio of cool to blue SG in the former, as previously discussed by \citet{Langer-Maeder95}.}. We take the luminosities of the stars in the LMC, and randomly select a fraction of those values to reflect the smaller sample size of the SMC. We then determine the most likely value of \lmax\ from this reduced sample, as well as the number of stars with \logl$>$5.4, $N_{5.4}$. We repeat this experiment $10^5$ times to determine the probability distributions of each of these quantities. We find that, if the two galaxies had the same intrinsic luminosity distribution, in the SMC we would expect an average \lmax=5.52$^{+0.05}_{-0.04}$, with $N_{5.4} = 2.6^{+1.4}_{-1.6}$ (67\% confidence limits). The probability of finding only one star with a luminosity above \logl=5.4 is 19\%, while the probability of zero stars above this limit is 4\%. Therefore, though not conclusive, the balance of probability suggests that \lmax\ is lower in the SMC than in the LMC\newtext{, a result which is the opposite to that predicted by evolutionary models. This is broadly in agreement with \citet{Humphreys83}, who found that the most luminous cool stars in their LMC and SMC samples were roughly the same, notwithstanding their selective samples and the issues that will be discussed in Sect.\ \ref{sec:previous}.}

Looking at the whole of the high-end of the luminosity distribution below \lmax, we can say that the evidence for the two galaxies having different intrinsic luminosity distributions is weak. A Kolmogorov-Smirnov test of the cumulative luminosity distributions for all stars with $5.0<$\logl$<$5.4 shows that there is a 60\% probability that the luminosities of the cool SGs in each galaxy are drawn from the same parent distribution. \newtext{Therefore, we do not find evidence to support the claim by \citet{Humphreys83} that the SMC has a steeper luminosity distribution than the LMC. }


\subsection{Comparisons with previous work} \label{sec:previous}

Our findings for \lmax\ are substantially lower than \logl=5.7-5.9 originally claimed in \citet{Humphreys-Davidson79}, later revised to \logl=5.66 by \citet{Humphreys83} and \citet{Elias85}. This is due in large part to a systematic downward revision of the luminosities of the cool supergiants in these galaxies. On average, we find luminosities that are 0.17dex fainter than those listed in Table 15 of Elias et al. The explanation for this is threefold; firstly, we are using a slightly lower distance modulus to the LMC (18.49, compared with 18.6 in \citealt{Humphreys79-LMC} and Elias et al.). Secondly, improvements in infrared photometry (higher sensitivity and spatial resolution, especially at longer wavelengths) means that we can obtain reliable spectral energy distributions for all stars in our sample, without relying on uncertain bolometric corrections derived from a subset of our sample. Thirdly, our treatment of extinction is fundamentally different to that of Elias et al. These authors inferred the total (interstellar + circumstellar) extinction by comparing the stars' colours to `intrinsic' colours of stars of the same spectral type. They did this by assuming that stars of the same spectral type have the same intrinsic $B-V$, regardless of metallicity, and used Galactic stars as templates. This involves de-reddening the Galactic stars, again accounting for both inter- and circumstellar extinction, which as Elias et al.\ themselves point out is extremely problematic. By contrast, we have used extinction maps to infer the foreground (interstellar) extinction, and assumed that the luminosity integrated between the $U$-band and 24\um\ is independent of circumstellar extinction. 

One caveat to our treatment of extinction is that the circumstellar dust could be clumpy, which would reduce the extinction per unit infrared-excess. However, this would cause us to overestimate the luminosity, further reducing \lmax, and increasing the disagreement with \citet{Elias85}. Further, it is unlikely that large amounts of cool dust, which emit at longer wavelengths than 24\um, are causing us to underestimate the luminosities of the stars in our sample. For all but a handful of stars the flux at 24\um\ is already negligible compared to that emitted at shorter wavelengths. For the star with the largest IR excess, WOH~G64 in the LMC, we have added in the 70\um\ {\it Spitzer}/MIPS photometry \citep{Jones17} to account for the contribution from cool dust. Even in this extreme case, the flux emitted between 24-70\um\ contributes only 0.05dex to the bolometric luminosity.

We can also compare to other estimates of the brightest cool SGs in the MCs. \citet[][ hereafter M09]{Massey09} revisited the luminosities of the RSGs in the LMC and SMC derived in L06, obtaining values of \lbol\ using both $V$-band and $K$-band photometry in conjunction with their bolometric corrections measured from MARCS model atmospheres. It was noted by M09 that the luminosities determined from the $V$-band were systematically higher than those determined at $K$, by an average of 0.12dex. For the stars we have in common with M09 we find good agreement between their $K$-based luminosities and our \lbol s. Therefore, we also reproduce the result that M09's $V$-band luminosities are brighter by $\sim$0.1dex. The explanation for this is compound. Firstly, the atmospheric models used by M09 (and references therein) are known to systematically overestimate the strengths of the TiO absorption bands at a given effective temperature \teff\ \citep{rsgteff}. This causes the total (foreground plus circumstellar) extinction and the \teff\ to both be underestimated.  This means that, for a {\it dereddened} $V$-band flux, one will overestimate the star's luminosity. Though the bolometric corrections in L06 seem consistent with ours (see Appendix), they are for {\it dereddened} $V$-band fluxes --  that is, the foreground {\it and} circumstellar extinction must first be accounted for. By contrast, our BCs already take into account the average circumstellar extinction at a given spectral type, and so  only knowledge of the foreground extinction is required.   

\begin{figure*}
\begin{center}
\includegraphics[width=7.5cm]{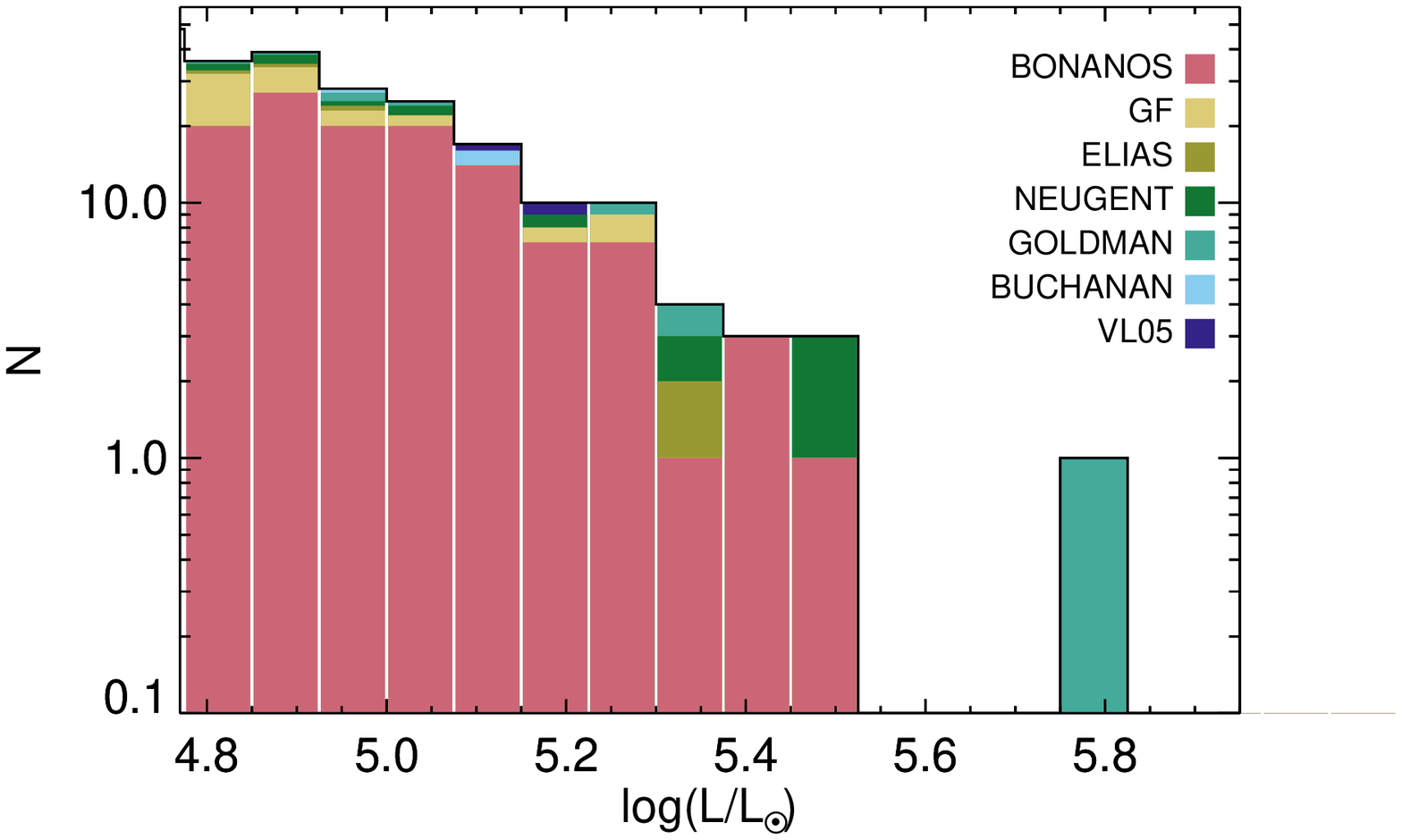}
\includegraphics[width=7.5cm]{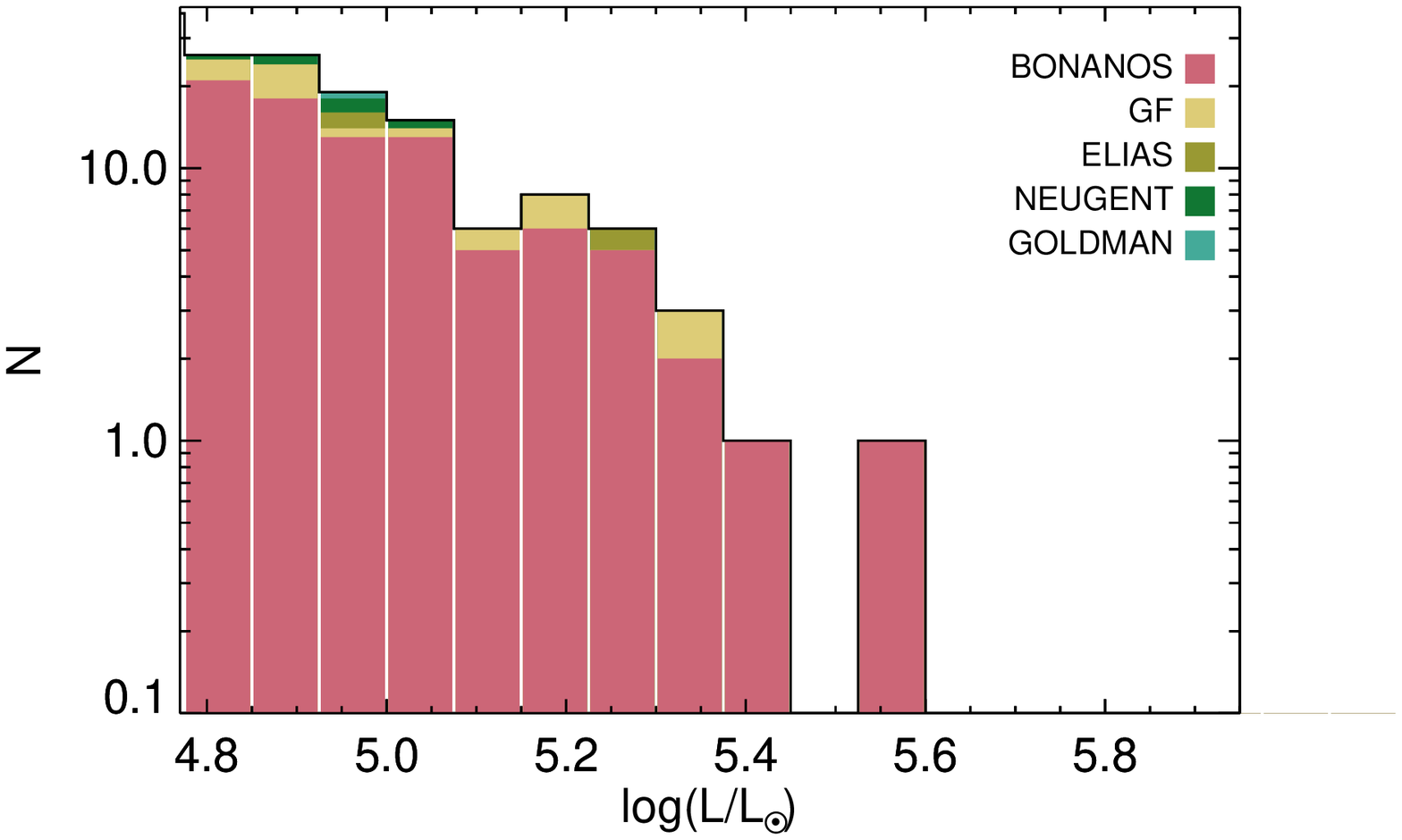}
\caption{Luminosity distributions for the LMC (left) and SMC (right), illustrating the contributions from the different samples included in this study.    }
\label{fig:compl}
\end{center}
\end{figure*}

\subsection{Statistical completeness}
\newtext{To assess the completeness of our samples of cool supergiants in the MCs, in \fig{fig:compl} we replot the luminosity distributions for each galaxy illustrating the contributions of the individual samples used in our study. The red bars in the figures show the luminosities of the objects in the Bonanos studies \citep{Bonanos09,Bonanos10}, while the yellow bars show those from \citet{Gonzalez-Fernandez15} which were {\it not} found in Bonanos. The subsequent colours, as shown in the legends, indicate the stars found in the corresponding survey that were not present in the other surveys listed above it in the legend. 

We see that almost all stars are found in the IR-based surveys of \citet{Gonzalez-Fernandez15}, \citet{Bonanos09,Bonanos10} and \citet{Neugent10,Neugent12}. The samples of dust-enshrouded stars \citep{vL05,Buchanan06,Goldman17,Goldman18} pick up a small number of luminous stars not detected in the IR surveys due to circumstellar extinction. All but a handful of stars from the selective survey of \citet[][ and references therein]{Elias85} are found in the systematic surveys listed earlier. 

The one object known to be luminous but not picked up in our catalogue search is IRAS~05280-6910. This object is part of the dense cluster NGC~1984, and is not well-resolved from the other stars in its parent cluster. This object was manually added in to our database, employing the high spatial resolution photometry of \citet{vanLoon05-clusters}}. 

From these results, we conclude that we are complete at high luminosities (\logl$\ga$5.0). Though the statistical completeness may begin to be non-negligible below this limit, this does not pose a problem for this current work since we are interested primarily in \lmax.


\begin{figure*}
\begin{center}
\includegraphics[width=7.5cm,bb=70 40 420 300,clip]{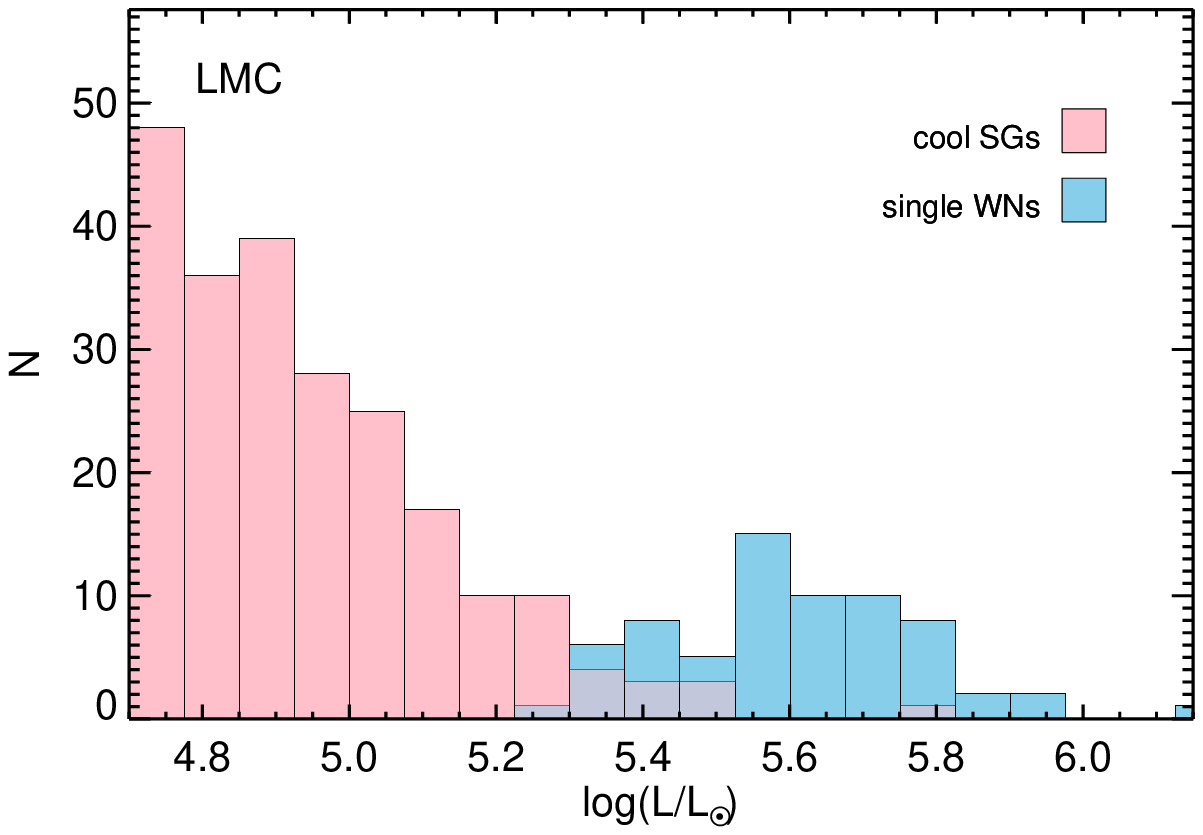}
\includegraphics[width=7.5cm,bb=70 40 420 300,clip]{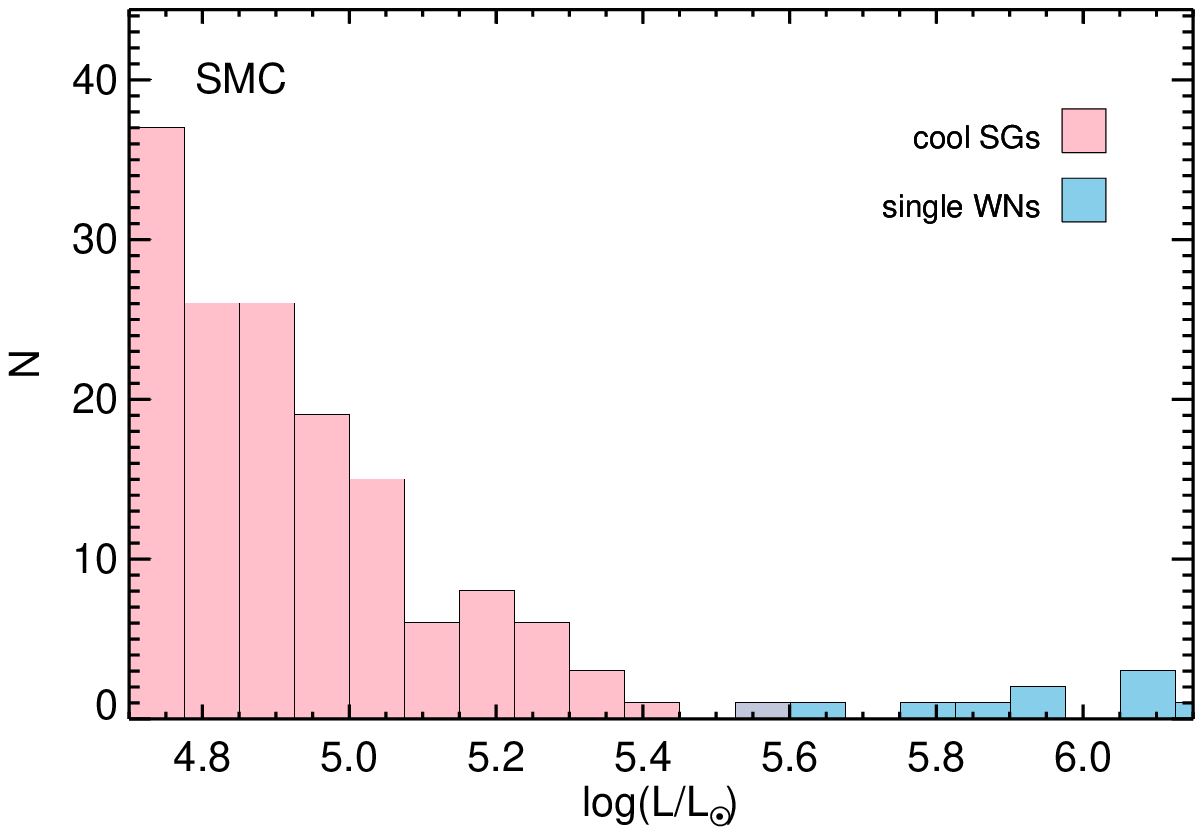}
\caption{Comparison of the luminosity distributions of RSGs and single WN stars in the two Magellanic Clouds.    }
\label{fig:WR}
\end{center}
\end{figure*}

\subsection{Comparison with Wolf-Rayet stars}
Since \lmax\ is thought to correspond to the initial mass at which single stars evolve directly from the MS to the WR phase, in \fig{fig:WR} we compare the cool SG luminosity distributions to those of nitrogen-rich WR (WN) stars. We choose to compare to WN stars as these are thought to be the least chemically evolved \citep{Crowther07}. For our sample of WNs, we have taken objects from \citet{Hainich14,Hainich15}, \citet{Neugent17} and \citet{Shenar16}, and discarded objects that were in known binaries in order to compare only those objects which result from single-star evolution. As a caveat however, we note that we are unable to rule out that some stars in this sample may still have experienced interaction with a companion which is no longer visible (either due to a merger or supernova). 

In the LMC (left panel of \fig{fig:WR}), the luminosity distributions splice together with a small overlap region between \logl=5.2-5.5 where presumably stars can experience a shortened cool SG phase before becoming a WR. In the SMC, there is no overlap between the two classifications of star, though this could easily be a result of lower number statistics. 

The results presented in \fig{fig:WR} are a clear demonstration of the commonly-held view that the evolution of a single star to the WR phase requires an initial mass above a certain threshold. \nntext{This threshold  roughly corresponds to the most massive cool SGs, with an overlap region where stars may possibly experience both a cool SG and WR phase} \citep[see also][]{vanLoon17}. Under the reasonable assumption that stars which evolve from the cool SG phase to the WR phase do so at $\sim$constant luminosity, the plot also indicates that the highest luminosity of a cool SG which will explode in that phase is \logl$\approx$5.2-5.3, i.e.\ the luminosity at which cool SGs and WRs co-exist. This is in agreement with the results of \citet{Davies-Beasor18}, who showed that, of all SNe with pre-explosion detections of the progenitor, the brightest (SN2009hd) had a pre-SN luminosity of \logl=$5.24\pm0.08$.

\subsection{Comparison to evolutionary models}
To make a quantitative comparison between our results and the expectations from evolutionary models, we perform a simple population synthesis analysis. We first generate a population of stars with masses drawn from a Salpeter initial mass function \citep[IMF, ][]{Salpeter55} and with ages sampled from a uniform random distribution between 0 and 50Myr, the latter being the expected lifetime of an 8\msun\ star. For each simulated star we interpolate an evolutionary track at that mass to determine its $L$ and effective temperature \teff\ at that star's age. If the age is greater than the star's lifetime, or the star is not in the cool supergiant region of the H-R diagram (i.e.\ \teff$>$7000K) then that star is discarded. To compare to our observations, we construct simulated luminosity distributions (LDs) with the same binning as the observed data, and renormalise the simulations to minimise the differences between the model and observed distributions in the range 4.7$\le$\logl$\le$5.2. We choose this luminosity range as here we expect to be largely complete whilst also having $\ga$5 stars per bin\footnote{If we are incomplete in this range, this would require the simulated LD to be moved upwards in number, which would also cause an increase in the simulated \lmax.}. 

The comparisons of the simulated and observed LDs are shown in \fig{fig:ldist}, the model predictions overplotted in blue. The models we have chosen are those of the Geneva group, who have published rotating and non-rotating models at Solar and SMC-like metallicity. At lower luminosities (\logl$\la$5.3), the slopes of the observed and simulated LDs match reasonably well, implying that the models correctly reproduce the relative numbers of cool SGs as a function of luminosity. The slope of this part of the LD is a combination of that of the IMF, the mass-luminosity relation for cool SGs (which may not be unique for a given mass), the mass-lifetime relation, and the fraction of this time that a star spends in the cool SG phase. Most evolutionary models agree that the ratio of the post-MS to MS lifetimes is $\sim$0.08-0.12, being at the lower end of this range for more massive stars, so one would not expect this to be a major source of uncertainty. \newtext{Therefore, assuming that star-formation in the MCs follows the standard Salpeter IMF, the similarity between the simulated and observed LD slopes indicates that the models are correctly predicting the convolution of the mass-luminosity relation and the cool SG lifetimes, at least in a relative sense\footnote{We note that we have made no attempt here to reproduce the absolute LDs of each galaxy by e.g.\ benchmarking against their global star formation rates. }}. Further, the result that the LDs look so similar between the two galaxies implies that this product is not strongly affected by metallicity. 

The discrepancy between the models and the observations comes when we look at the highest luminosities. In the LMC (left panel of \fig{fig:ldist}) the Solar metallicity models provide a very good match to the observed LD: at \logl$\ga$5.6, we see a decrease in the fraction of time spent by stars in the cool SG phase. This causes a downturn in the predicted LD, fitting the observed \lmax\ to within the errors. However, models with an LMC-like metallicity (not available at the time of writing) would have a larger \lmax, owing to the reduced mass-loss rates on the MS as discussed earlier. The effect is more pronounced when we look at the SMC (right panel of \fig{fig:ldist}). Both the rotating and non-rotating models predict that we should be seeing cool SGs with luminosities of \logl=5.7-5.8. Quantitatively, we see only one star in the SMC with a luminosity above \logl$\ge$5.36, compared to 25 (18) predicted by rotating (non-rotating) models\footnote{We note that rotational velocities as high as those of the rotating Geneva models are rarely seen in the LMC \citep{Ramirez-Agudelo13}.}. This implies that, at higher luminosities, either stars cannot evolve to the cool SG phase {\it or} this phase is so short that it is unlikely to be observable. In either case, this result implies that the latest population synthesis models \citep[e.g.][]{Leitherer14} are underestimating the ionising fluxes and producing integrated colours that are too red for populations at low metallicity.

\subsection{Possible causes of a reduced \lmax}
Having argued in the previous section that single-star models substantially over-predict the numbers of high luminosity cool SGs, we now discuss various potential solutions to this discrepancy. For the purposes of this discussion we characterize this discrepancy in terms of the apparent \lmax, that is the brightest cool SG that one is likely to detect in a finite population of stars. 

The first obvious aspect to discuss is that of stellar winds. It is commonly argued that the upper luminosity limit for RSGs is a result of higher mass stars having stronger winds, losing a larger fraction of their initial mass during their lifetimes. Above some mass, almost all the H-rich portion of envelope is lost prior to the RSG phase, keeping the star in the blue. Hence, increasing the mass-loss rate \mdot\ would in turn reduce the observed \lmax. 

On the main-sequence, the mass-loss rate prescriptions of  O~stars employed in the Geneva models seem to be supported by observations \citep{Mokiem07}, and so there is little justification in increasing these. Further, we have shown here that \lmax\ does {\it not} increase with decreasing metallicity, as one would expect if line-driven winds were the cause. This suggests that, if stellar winds govern the observed value of \lmax, then these winds would have to be metallicity-independent continuum-driven winds, such as those suggested for Luminous Blue Variables \citep[e.g.][]{Smith-Owocki06}.

\begin{figure*}
\begin{center}
\includegraphics[width=7.5cm]{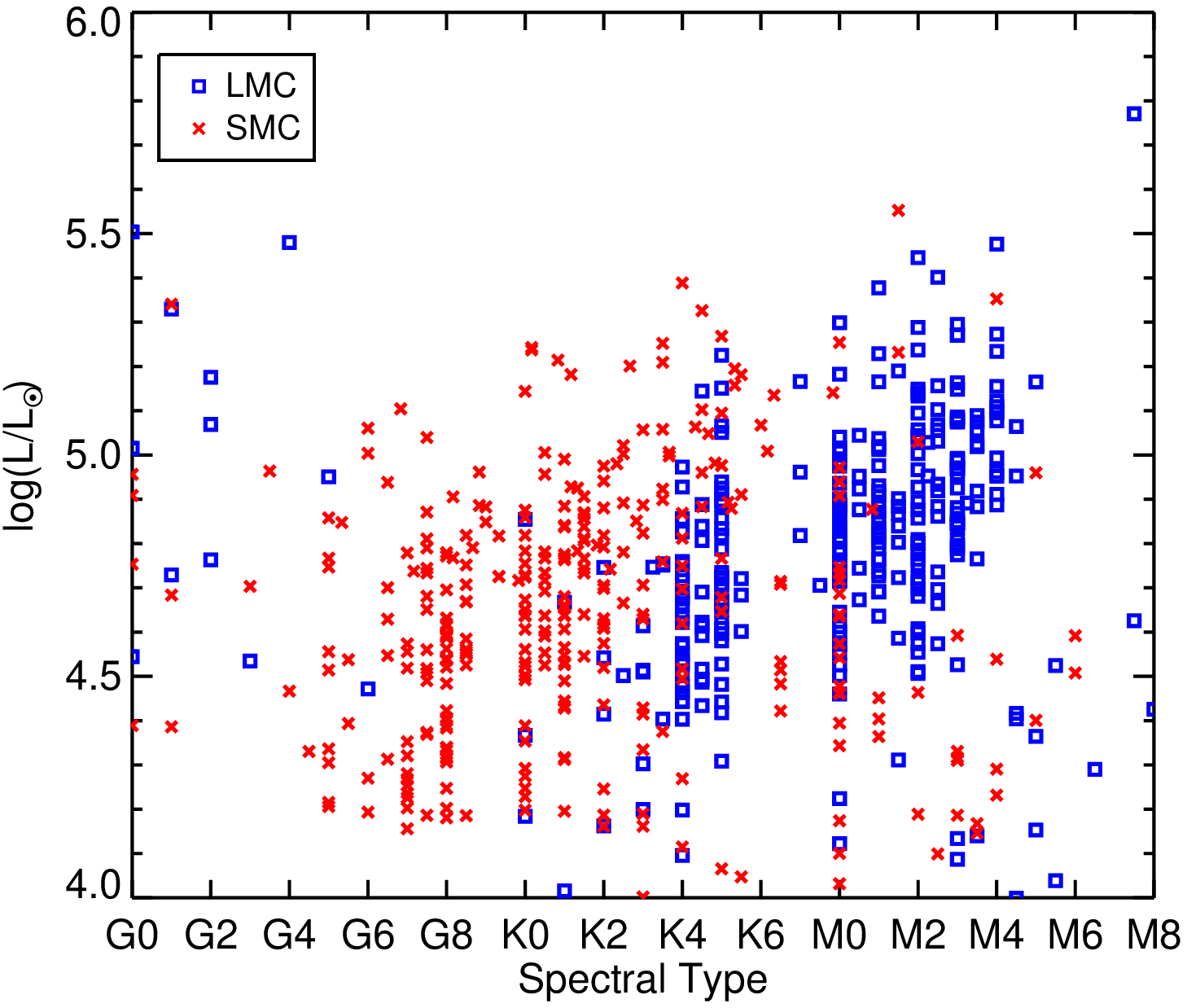}
\includegraphics[width=7.5cm]{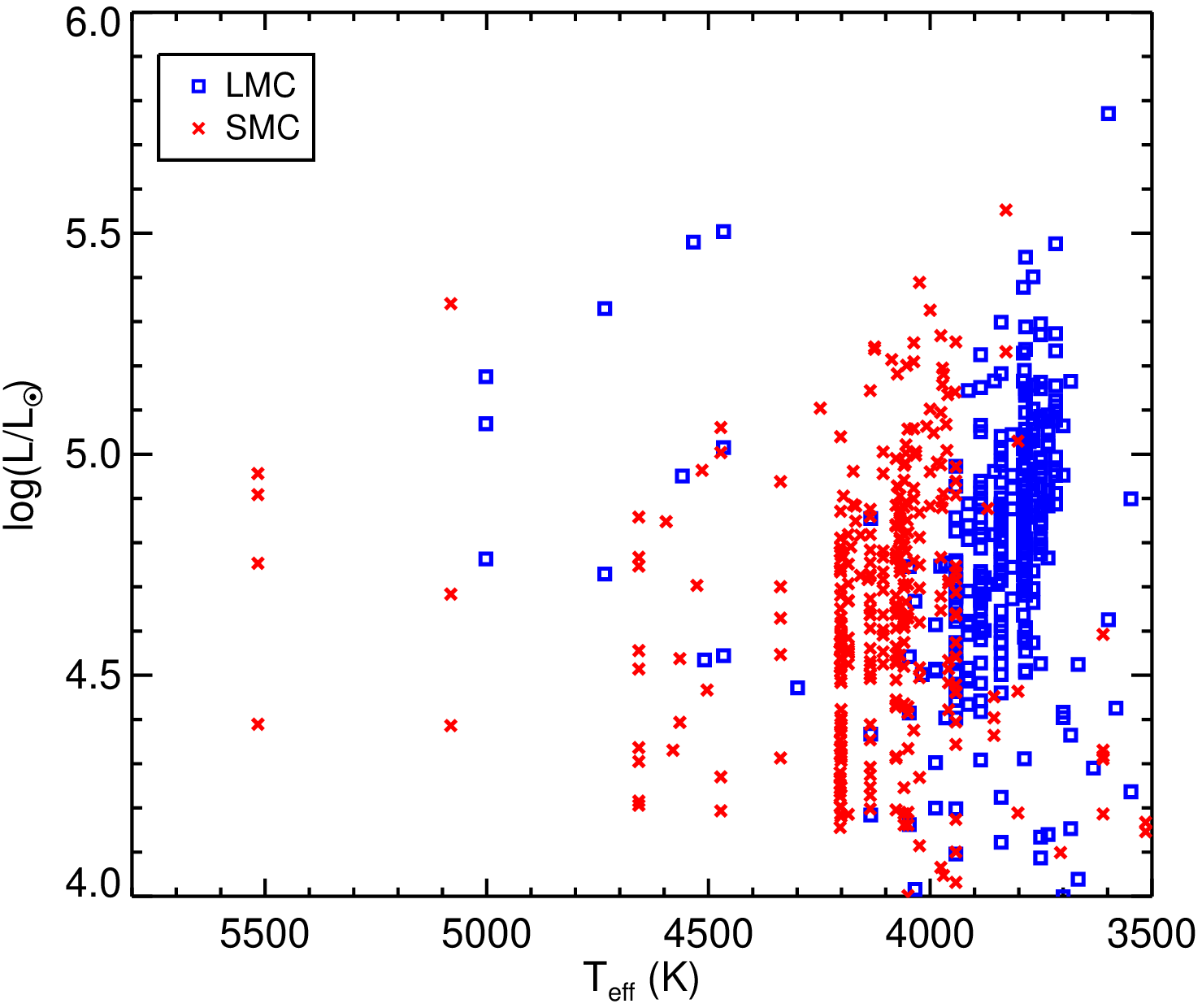}
\caption{{\it Left:} Observational H-R diagram for the cool SGs in each galaxy.
{\it Right:} Same as the left panel, but with spectral types converted to effective temperatures using the temperature scales of \citet{Tabernero18}. }
\label{fig:hrd}
\end{center}
\end{figure*}

In terms of mass-loss during the cool SG phase, increasing \mdot\ for RSGs seems poorly justified. Indeed, measurements of how \mdot\ evolves throughout the RSG phase suggest that all evolutionary models are currently {\it overestimating} the total integrated mass lost during this time, at least for stars with initial masses $\sim$16\msun\ \citep{Beasor-Davies18}. This would result in the opposite of what we see in the MCs: it would make higher mass single stars more likely to explode in the RSG phase, moving \lmax\ to higher luminosities. \newtext{One way out of this would be if stars experience a short period of enhanced mass-loss towards the end of the RSG phase which is so brief that only a few stars per galaxy would be in this phase at any one time. Such stars may appear as OH/IR stars, characterized by large infrared excesses and circumstellar maser emission. If we concentrate on the LMC, there are 4 out of 73 stars in our sample which have OH masers with luminosities \logl$\ge$5.0 (above which we consider our sample to be complete). If all RSGs experience an OH/IR phase, for a canonical RSG lifetime of $10^6$\,yrs this suggests an OH/IR lifetime of a few $\times 10^4$\,yrs. With a typical mass-loss rate of  $\sim 10^{-4}$\,\msunyr\ \citep[e.g.][]{Goldman17}, there is the potential to lose several Solar masses of envelope during this phase, which would dwarf that lost up to that point in the star's life. }


One other way to reduce \lmax\ would be to invoke the effects of binary mass transfer as a way to increase the amount of mass lost during a star's life. Specifically, a trend of increasing the interacting binary fraction with initial stellar mass would have the effect of reducing the probability of forming cool SGs at high luminosities. \citet{Moe-diStefano17} argue that the single-star fraction decreases as a function of initial stellar mass, whilst the companion frequency at short periods increases. Together, these effects would serve to decrease the likelihood of a primary becoming a cool SG at higher initial masses. 

Quantitative modelling of the luminosity distribution of cool SGs for a population of stars would need to account for the IMF and star-formation rate, as well as the mass-dependence of the lifetimes of the MS and post-MS phases, the luminosity evolution within the cool SG phase, and the binary fractions and period distributions. Such work is beyond the scope of this present study and will be the subject of a future paper.

\subsection{Observational H-R diagram}
To compare the differences in the cool supergiant populations of the two MCs, in \fig{fig:hrd} we plot an observational H-R diagram of all stars in our two samples. \newtext{ In the left-hand panel we plot spectral type on the horizontal axis; on the right we plot the same but employing the temperature scale of \citet{Tabernero18}}. Though incompleteness effects are obvious below \logl$\la$4.7, one can clearly see that there is an offset in spectral types between the two galaxies.  

The shift to earlier average spectral types of RSGs from early-M to late-K as one moves from the LMC to the SMC is well-known \citep[e.g.][ and references therein]{Elias85}. \newtext{Up until recently there has been little compelling evidence} as to whether this represents a shift of the Hayashi limit to higher \teff\ at lower metallicity, or whether it is simply an effect of lower metal abundances reducing the strengths of the TiO absorption lines which define the transition from K to M types \citep{rsgteff}. However, what \fig{fig:hrd} shows is that the systematic shift to earlier spectral types in the SMC compared with the LMC goes beyond a shift from M to K, but also reaches to late G-types. There are very few cool SGs in the LMC with spectral types earlier than K2. By contrast, there are many cool SGs in the SMC with spectral types earlier than K. This {\it cannot} be explained as a metallicity effect alone, as the differences between G and K classifications are driven mainly by ionisation rather than simply strengths of lines. This difference must then be caused by a metallicity dependence of (a) the temperature of the Hayashi limit, and/or (b) the speed at which supergiants cross the H-R diagram\newtext{, either on their way {\it to} or back {\it from} the RSG phase. } \nntext{Interestingly, there {\it are} stars in the SMC which have late spectral types ($>$M2), particularly at lower luminosities. Many of these objects could be super-AGB stars, the descendants of intermediate mass stars.  }

\newtext{ In the right-panel of \fig{fig:hrd} we convert the spectral-types of the stars to effective temperature using the calibration of \citet{Tabernero18}, based on comparisons to LTE model atmospheres. We have many stars in common with Tabernero et al, which is based on the \citet{Gonzalez-Fernandez15} sample, though our sample has a higher level of completeness particularly for objects with large reddening. Predictably, our results show the same as Tabernero et al., specifically that the LMC stars are systematically cooler than those in the SMC. We note that in the Tabernero scale, the dispersion in \teff\ at a given spectral type is quite large, up to $\pm$100K, with a 200K spread being as large as the \teff\ difference between e.g. K2 and M1 spectral types. This explains the apparent paradox whereby no temperature scale was detected by \citet{MCpaper} yet object-by-object comparisons with Tabernero et al.\ showed good agreement; the Davies et al.\ sample of $\sim$10 objects per galaxy was too small to detect the subtle variations in \teff\ as a function of spectral type. }


\section{Conclusions} \label{sec:conc}
We have combined various surveys of cool supergiants and used multi-wavelength survey photometry from the $U$-band to the mid-infrared to redetermine the luminosity distributions of cool massive stars in the Large and Small Magellanic Clouds. Our main findings are as follows:

\begin{itemize}
\item The most luminous cool stars in the LMC and SMC have \logl=5.77 and 5.55 respectively, though the brightest of these is highly variable. The next most luminous stars have \logl=5.50 and 5.36 respectively. If these stars represent the upper luminosity limit \lmax\ (otherwise known as the Humphreys-Davidson limit), this is a downward revision of the previously quoted limit of \logl$\simeq$5.7 in the literature. 
\item We find no evidence to support the commonly-held view that \lmax\ is higher at lower metallicity. Indeed our results indicate that it is unlikely that \lmax\ in the SMC is higher than in the LMC, even after accounting for low number statistics. This argues against metallicity-dependent mass-loss being the cause of \lmax. 
\item A population synthesis analysis of the two luminosity distributions reveals that the Geneva evolutionary models predict too many luminous cool stars, particularly in the SMC. Specifically, models predict $>$19 cool supergiants in the SMC with luminosities \logl$>$5.36, whereas we see only one.  
\item The luminosity distributions of cool supergiants splice together with those of apparently-single Wolf-Rayet stars in each of the MCs, suggesting a changing evolutionary sequence of massive stars with increasing initial mass. 
\item The spectral types of cool supergiants are earlier in the SMC than in the LMC, a well-known result. However, the shift extends beyond that of M to K, with a substantial number of G supergiants in the SMC. This implies that the average temperatures of cool supergiants are hotter at lower metallicities. 
\end{itemize}


\section*{Acknowledgements}
The authors would like to thank the referee Jacco van Loon for suggestions and comments which helped us improve the paper, Fabian Schneider for enquiring about the upper luminosity of cool supergiants in the Large Magellanic Cloud, and Nathan Smith for useful comments and discussion. This publication makes use of data products from the Wide-field Infrared Survey Explorer, which is a joint project of the University of California, Los Angeles, and the Jet Propulsion Laboratory/California Institute of Technology, funded by the National Aeronautics and Space Administration. This publication makes use of data products from the Two Micron All Sky Survey, which is a joint project of the University of Massachusetts and the Infrared Processing and Analysis Center/California Institute of Technology, funded by the National Aeronautics and Space Administration and the National Science Foundation. This work made use of the IDL astronomy library, available at {\tt https://idlastro.gsfc.nasa.gov}, and the Coyote IDL graphics library.




\bibliographystyle{mnras}
\bibliography{/Users/astbdavi/Google_Drive/drafts/biblio} 



\appendix

\begin{figure*}
\begin{center}
\includegraphics[width=15cm]{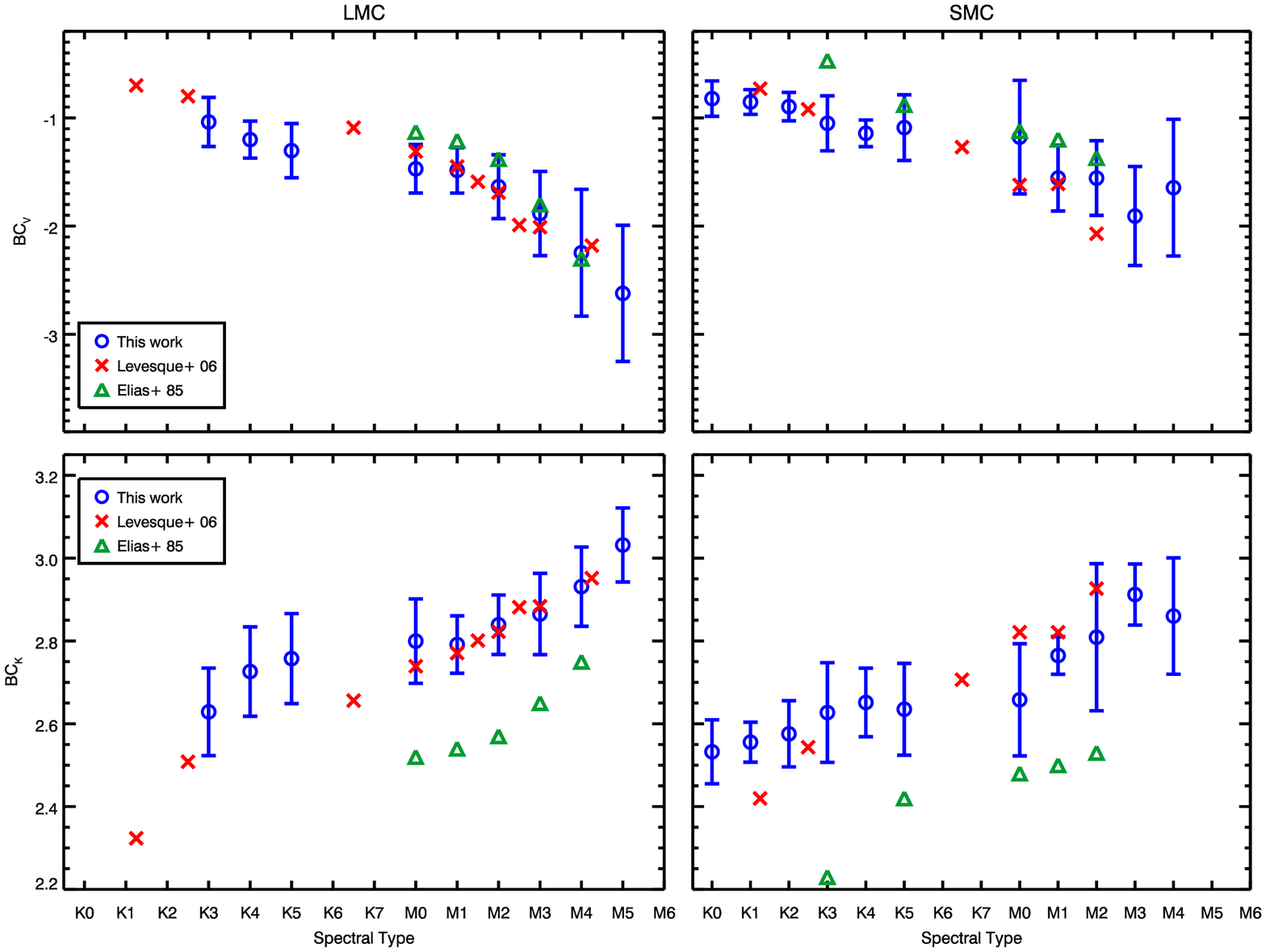}
\caption{Bolometric corrections as a function of spectral type in the two Magellanic Clouds, compared to those measured by \citet{Elias85} and \citet{Levesque06}.    }
\label{fig:BCs}
\end{center}
\end{figure*}

\section{Bolometric corrections of cool supergiants in the Magellanic Clouds}
Since we now have bolometric luminosities for each star in our sample, we can derive empirical bolometric corrections (BCs) as a function of spectral type. To do this, we take each star's photometry at Johnson-$V$, DENIS-$I$ and 2MASS-$K_S$, dereddened according to the extinction law of \citet{Gordon03}. We do {\it not} remove any circumstellar component to the total extinction. Therefore, the BCs we provide should be applied to photometry {\it without} attempting to compensate for circumstellar extinction, which is notoriously difficult to estimate given its degeneracy with foreground extinction and the \teff\ of the star. Our BCs already account for an average amount of circumstellar extinction for stars of the same spectral type and metallicity of the LMC and SMC. 

At each spectral subtype, we take the average BC to be the median of all stars within $\pm$0.5 subtypes. We have not estimated the BC at any subtype where we had less than 5 stars. We define the error at each subtype to be the standard deviation of stars in that bin within 2.5$\sigma$ of the mean. This reduces the impact of the small number of outliers, typically caused by poor photometry. \newtext{However, we note that there are some statistical outliers beyond these limits, particularly at later types.  }

In \fig{fig:BCs} we compare our BCs at $V$ and $K$ to those of \citet{Elias85} and L06. The BCs in Elias et al.\ were empirically derived, whereas those in L06 were determined from atmospheric models. Whilst at face value there appears to be good agreement at $V$ between all three studies and at $K$ between this work and L06,  one must keep in mind that the BCs of Elias et al.\ and Levesque et al.\ were defined for {\it unreddened} photometry. Therefore, removing a circumstellar component to the extinction of e.g.\ $A_V \simeq 0.5$ prior to applying the BC would result in overestimating the luminosity from the $V$-band photometry by $\sim$0.2dex. 

\begin{table}
\caption{Average bolometric corrections as a function of spectral type for the two Magellanic Clouds. The filter systems are Johnson $V$, DENIS-$I$ and 2MASS $K_s$.}
\begin{center}
\begin{tabular}{lccc}
\hline \hline
SpT & \bcv & \bci & \bck \\
\hline
{\it LMC} \\
K0-K3 & $-$1.15 $\pm$  0.23 &  0.50 $\pm$  0.07 &  2.69 $\pm$  0.11 \\ 
   K4 & $-$1.16 $\pm$  0.17 &  0.51 $\pm$  0.07 &  2.69 $\pm$  0.11 \\ 
K5-K6 & $-$1.19 $\pm$  0.25 &  0.51 $\pm$  0.06 &  2.70 $\pm$  0.11 \\ 
K7-M0 & $-$1.43 $\pm$  0.22 &  0.46 $\pm$  0.10 &  2.77 $\pm$  0.10 \\ 
   M1 & $-$1.56 $\pm$  0.21 &  0.42 $\pm$  0.10 &  2.81 $\pm$  0.07 \\ 
   M2 & $-$1.72 $\pm$  0.29 &  0.37 $\pm$  0.14 &  2.85 $\pm$  0.07 \\ 
   M3 & $-$1.91 $\pm$  0.39 &  0.31 $\pm$  0.19 &  2.89 $\pm$  0.10 \\ 
   M4 & $-$2.12 $\pm$  0.59 &  0.24 $\pm$  0.20 &  2.94 $\pm$  0.10 \\ 
   M5 & $-$2.36 $\pm$  0.63 &  0.16 $\pm$  0.17 &  3.00 $\pm$  0.09 \\ \smallskip \\ 
{\it SMC} \\
  G5 & $-$0.51 $\pm$  0.27 &  0.75 $\pm$  0.04 &  2.37 $\pm$  0.24 \\ 
   G6 & $-$0.57 $\pm$  0.26 &  0.73 $\pm$  0.08 &  2.40 $\pm$  0.11 \\ 
   G7 & $-$0.62 $\pm$  0.20 &  0.71 $\pm$  0.07 &  2.43 $\pm$  0.09 \\ 
   G8 & $-$0.68 $\pm$  0.10 &  0.69 $\pm$  0.05 &  2.46 $\pm$  0.06 \\ 
   G9 & $-$0.75 $\pm$  0.10 &  0.67 $\pm$  0.03 &  2.49 $\pm$  0.05 \\ 
   K0 & $-$0.81 $\pm$  0.16 &  0.65 $\pm$  0.05 &  2.52 $\pm$  0.08 \\ 
   K1 & $-$0.88 $\pm$  0.11 &  0.63 $\pm$  0.06 &  2.55 $\pm$  0.05 \\ 
   K2 & $-$0.95 $\pm$  0.13 &  0.62 $\pm$  0.07 &  2.58 $\pm$  0.08 \\ 
   K3 & $-$1.02 $\pm$  0.25 &  0.60 $\pm$  0.08 &  2.61 $\pm$  0.12 \\ 
   K4 & $-$1.10 $\pm$  0.12 &  0.59 $\pm$  0.12 &  2.64 $\pm$  0.08 \\ 
K5-K6 & $-$1.18 $\pm$  0.30 &  0.57 $\pm$  0.12 &  2.67 $\pm$  0.11 \\ 
K7-M0 & $-$1.43 $\pm$  0.52 &  0.53 $\pm$  0.11 &  2.76 $\pm$  0.14 \\ 
   M1 & $-$1.51 $\pm$  0.30 &  0.52 $\pm$  0.12 &  2.79 $\pm$  0.05 \\ 
   M2 & $-$1.60 $\pm$  0.34 &  0.51 $\pm$  0.24 &  2.82 $\pm$  0.18 \\ 
   M3 & $-$1.70 $\pm$  0.46 &  0.50 $\pm$  0.19 &  2.85 $\pm$  0.07 \\ 
   M4 & $-$1.79 $\pm$  0.63 &  0.49 $\pm$  0.15 &  2.88 $\pm$  0.14 \\ 
   \hline
\end{tabular}
\end{center}
\label{tab:bcs}
\end{table}%



\bsp	
\label{lastpage}
\end{document}